\documentclass[twocolumn,secnumarabic,amssymb,nobibnotes,aps,prb]{revtex4-1}
\usepackage{color}
\usepackage{graphicx}
\usepackage{natbib}
\usepackage{url}
\usepackage{textcase}
\usepackage{bm}
\usepackage{amsmath} 
\usepackage{ulem} 

\newcommand{\kCl}{$\kappa$-Cl}
\newcommand{\kBr}{$\kappa$-Br}
\newcommand{\CuNCS}{$\kappa$-CuNCS}

\begin{document}

\title[Origin of the glass-like dynamics in molecular metals $\kappa$-(BEDT-TTF)$_2$X]{Origin of the glass-like dynamics in molecular metals $\kappa$-(BEDT-TTF)$_2$X: implications from fluctuation spectroscopy and \textit{ab initio} calculations}

\author{Jens M\"uller}
\email[Email: ]{j.mueller@physik.uni-frankfurt.de}
\author{Benedikt Hartmann}
\author{Robert Rommel}
\author{Jens Brandenburg}
\author{Stephen M. Winter}
\affiliation{Institute of Physics, Goethe-University Frankfurt, 60438 Frankfurt (M), SFB/TR49, Germany}
\author{John A. Schlueter}
\altaffiliation[Present address: ]{Division of Materials Research, National Science Foundation, Arlington, VA 22230, USA}
\affiliation{Argonne National Laboratory, Materials Science Division, Argonne, IL 60439, USA}

\date{\today}

\begin{abstract}
We have studied the low-frequency dynamics of the charge carriers in different organic charge-transfer salts $\kappa$-(BEDT-TTF)$_2$X with polymeric anions X by using resistance noise spectroscopy. Our aim is to investigate the structural, glass-like transition caused by the conformational degrees of freedom of the BEDT-TTF molecules' terminal ethylene groups. 
Although of fundamental importance for studies of the electronic ground-state properties, 
the phenomenology of the glassy dynamics is only scarcely investigated and its origin is not understood. Our systematic studies of fluctuation spectroscopy of various different compounds reveal a universal, pronounced maximum in the resistance noise power spectral density related to the glass transition.
The energy scale of this precess can be identified with the activation energy of the glass-like ethylene endgroup structural dynamics as determined from thermodynamic and NMR measurements. For the first time for this class of 'plastic crystals', we report a typical glassy property of the relaxation time, namely a Vogel-Fulcher-Tammann law, and are able to determine the degree of fragility of the glassy system. Supporting {\it ab initio} calculations provide an explanation for the origin and phenomenology of the glassy dynamics in different systems in terms of a simple two-level model, where the relevant energy scales are determined by the coupling of the ethylene endgroups to the anions.
%
\end{abstract}

\pacs{74.70.Kn, 71.30.+h, 72.70.+m}

\maketitle
\section{Introduction}
\begin{figure}[tbp]
\begin{center}
\includegraphics[width=\columnwidth]{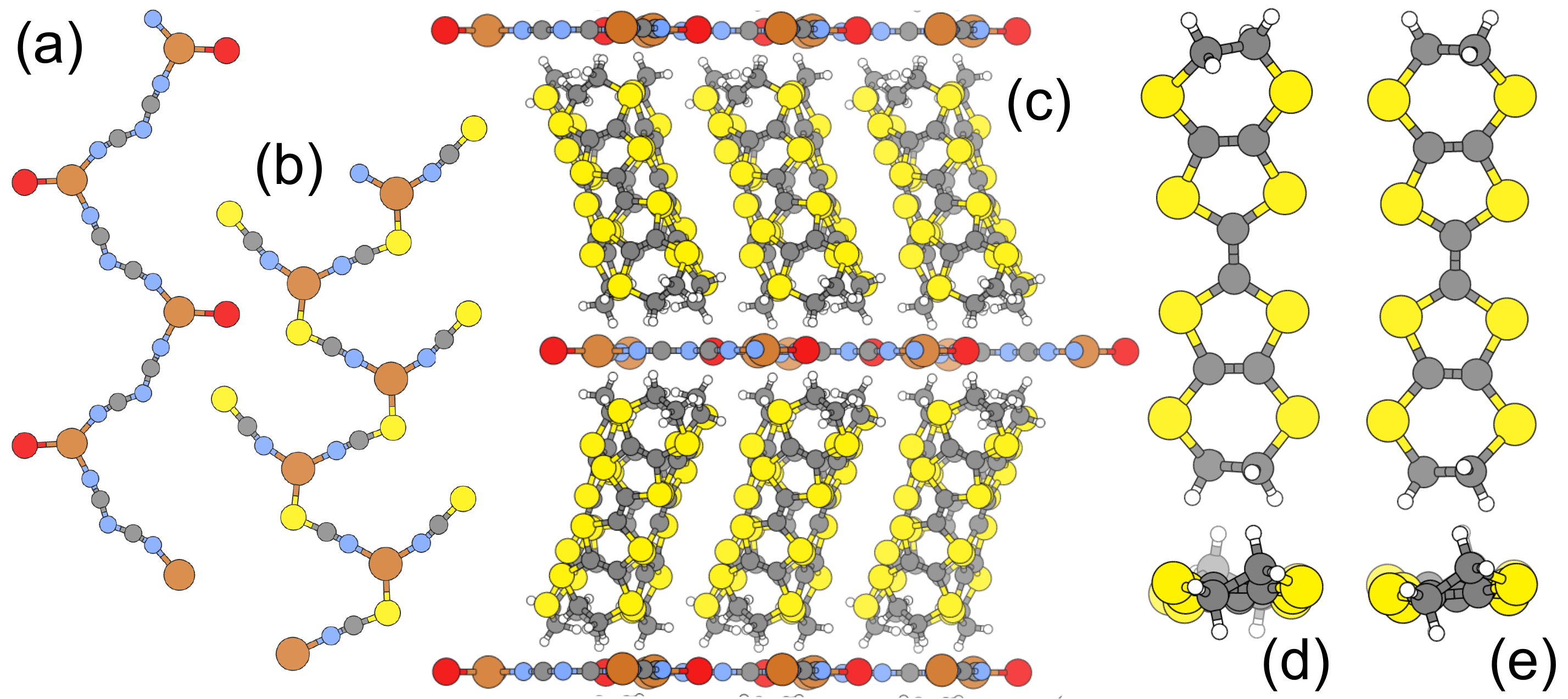}
\caption{\small{(color online) (a), (b) 1D coordination polymer anions Cu[N(CN)$_2$]Br and Cu(SCN)$_2$, respectively. (c) Crystal structure of $\kappa$-(ET)$_2$Cu[N(CN)$_2$]Br, side view showing the sequence of conducting ET layers and insulating anion sheets. (d), (e) Schematic view of the relative orientations of the ET molecules' ethylene endgroups (EEG) in the staggered (S) and eclipsed conformation (E), respectively.}}
\label{struktur}
\end{center}
\end{figure}

In the last years, the quasi-two-dimensional (quasi-2D) organic charge-transfer salts (BEDT-TTF)$_2$X, where BEDT-TTF (bis-ethylenedithio-tetrathiafulvalene, commonly abbreviated as ET) represents C$_6$S$_8$[C$_2$H$_4$]$_2$, have attracted considerable attention as model systems for studying the physics of correlated electrons  in reduced dimensions\cite{Ishiguro1998,Toyota2007,Lebed2008,Powell2011}. The observed rich phenomenology of ground states is due to both the highly tunable nature of the correlation strength of the charge carriers and the strong coupling of the latter to intra- and intermolecular vibrational modes of the underlying crystal lattice. The materials are highly anisotropic systems with alternating conducting and insulating layers, where the former contains the electron-donor ET molecules, for which the overlap of the molecular $\pi$-orbitals of adjacent molecules leads to the formation of a quasi-2D conduction band.
%
The $\kappa$-phase organic conductors $\kappa$-(ET)$_2$X with polymeric anions X$^-$ are the most intensively studied systems of this family because of their superconducting, magnetic and dielectric properties, and in view of the current debate related to the Mott metal-insulator transition (MIT) \cite{Kagawa2009,Zacharias2012,Lunkenheimer2012,Furukawa2015}. The latter can be accessed by varying the ratio of bandwidth $W$ to effective onsite Coulomb repulsion $U$ by means of physical or chemically-induced pressure\cite{Kanoda1997}. For example, the ground state of the antiferromagnetic Mott insulator $\kappa$-(ET)$_2$Cu[N(CN)$_2$]Cl ($T_N = 27$\,K, denoted as $\kappa$-Cl hereafter) can be turned into a superconducting state by either applying a moderate pressure of $\sim 300$\,bar or by modifying the anion X: $\kappa$-(ET)$_2$Cu[N(CN)$_2$]Br ($\kappa$-Br)  and $\kappa$-(ET)$_2$Cu(NCS)$_2$ ($\kappa$-CuNCS)  are ambient-pressure superconductors with $T_c = 11.6$\,K and 10.4\,K, respectively. Likewise, stepwise deuteration, {\it i.e.}, replacing hydrogen by deuterium in the terminal ethylene moieties [C$_2$H$_4$] of the ET molecules, drives the system $\kappa$-Br from the superconducting towards the Mott insulating phase \cite{Kawamoto1997}. 
%
%
In the past years, the intrinsic structural disorder associated with the thermal dynamics of the terminal ethylene groups has been found to play an important role for the ground-state properties of the strongly correlated electrons.
As shown schematically in Fig.\,\ref{struktur}, the [C$_2$H$_4$] ethylene endgroups (EEG hereafter) can adopt two different conformations, {\it i.e.}, when viewed along the central C=C bond of the ET molecule, their orientation is either eclipsed (E) or staggered (S). For the $\kappa$-(ET)$_2$X salts discussed in this work, the population of the E and S states is thermally disordered at room temperature due to strong thermal vibrations. Upon cooling to low temperatures, the EEG tend to adopt one of the two possible conformations, depending on the anion and the crystal structure, {\it i.e.}, E for and \kCl\ and \kBr\ and S for \CuNCS\ \cite{Ishiguro1998}.
However, in these three salts, the ordering of the EEG at low temperatures cannot be completed for kinetic reasons, since upon cooling through a characteristic temperature $T_g$ their molecular motion slows down so rapidly that thermodynamic equilibrium cannot be reached. At this so-called {\it glass-like transition}, a short-range structural order becomes frozen-in. The glass temperature $T_g$ depends in a characteristic way on the cooling rate $q_c$, {\it i.e.}, faster cooling rates result in a higher $T_g$ and therefore a larger degree of quenched disorder \cite{Mueller2002,Hartmann2014}.
Below this temperature, the thermal vibrations are frozen and no longer contribute to the specific heat and thermal expansion \cite{Donth2001}. Thus, for the present materials, the glassy characteristics of the ordering transition becomes evident in thermodynamic quantities, as frequency-dependent specific heat \cite{Akutsu2000} and most pronounced in large anomalies in the anisotropic, linear coefficients of thermal expansion \cite{Mueller2002,Mueller2004} exhibiting thermal hysteresis and  under- and overshoot behavior in the warming curves. 
For thermal expansion measurements, a possible way to determine the glass-transition temperature $T_g$ is to consider the intersection of warming and cooling curves taken at the same rates $|q|$, which results in $-|q| \cdot {\rm d}\tau/{\rm d}T|_{T_g} \cong 1$, where $\tau$ is the relaxation time of the corresponding molecular entities, see Refs.\ \cite{Mueller2002,Mueller2004}. This analysis yields $T_g = 72$\,K for $\kappa$-Cl and $T_g = 75$\,K for $\kappa$-Br. $\kappa$-CuNCS exhibits a more complex behavior, namely a sequence of two transitions at $T_{g_1} = 70$\,K and $T_{g_2} = 53$\,K. 
In contrast, no evidence of a distinct glass transition has been observed in other salts, such as the quantum spin liquid candidate $\kappa$-(ET)$_2$Cu$_2$(CN)$_3$ 
(denoted as $\kappa$-Cu$_2$(CN)$_3$) \cite{deSouza2015}, despite thermal disorder of the EEGs in this material at high temperatures. In this article we aim to provide an explanation for these differences.

%
From a fundamental point of view, the nature of the glass (or glass-like \cite{comment1}) transition is one of the most intriguing and still unresolved problems in condensed matter physics \cite{Anderson1995,Ediger1996,Debenedetti2001}. The phenomenon of glass formation occurs in a variety of different materials \cite{Angell1988,Boehmer1993,Angell1995}, whereas the tendency of building a glass former depends on the complexity of the contributing molecules. Besides ac specific heat, thermal expansivity and compressibility measurements \cite{Angell1976,Birge1986,Dixon1990}, it is important to extract information of the frequency- and temperature-dependent dynamics of the glass building processes. To that end, a number of sophisticated experimental techniques like viscosity measurements for liquids \cite{Angell1988} and broadband dielectric spectroscopy of solids \cite{Lunkenheimer2000} have been established. In particular the latter technique has become a powerful tool for studying the glassy dynamics of 'plastic crystals' (or orientational glasses) \cite{Hoechli1990}, a category to which the present materials belong, since they are characterized by disorder with respect to the orientational degrees of freedom of the translationally ordered molecules \cite{Lunkenheimer2000}. Dielectric spectroscopy has been successfully applied to many quasi-1D \cite{Nad2006,Staresinic2006} and quasi-2D organic charge transfer salts \cite{Pinteric1999,AbdelJawad2010,Ivek2010,Lunkenheimer2012,Iguchi2013,Pinteric2014}, where both short-range relaxor-type as well as long-range ferroelectric order has been observed, mostly in charge-ordered and insulating phases. The glassy dynamics of the EEG rotational degrees of freedom, however, has not been investigated by dielectric spectroscopy, since even the conductivity of semiconducting \kCl\ is too high for temperatures $T > T_g$ \cite{Lunkenheimer2012}.\\ 
In previous studies we were able to show that resistance $1/f$-noise spectroscopy provides a useful alternative method for studying glassy molecular dynamics in conducting materials, where dielectric spectroscopy is difficult to apply \cite{JMueller2009b,JMueller2011}. A prerequisite is a sufficiently large coupling of the vibrational properties of the glass-forming entities to the electric conductivity and its fluctuations, which recently has been directly observed in NMR experiments \cite{Kuwata2011}. The specific effects of this large coupling is twofold:\\
(i) 
The anisotropic change of in-plane lattice parameters at $T_g$ \cite{Mueller2002} results in a smaller average ratio $t_{\rm inter}/t_{\rm intra}$ of the inter- and intra-dimer transfer integrals for more rapid cooling, which in turn leads to a smaller effective $W/U$ \cite{Sasaki2005}, {\it i.e.}, the strength of electronic correlations can be controlled by varying the degree of frozen-in disorder at $T_g$. In a recent work, we could demonstrate that, when thermally coupled to a low-temperature heat bath, a pulsed heating current through the sample causes a very fast relaxation with cooling rates at $T_g$ of the order of several 1000\,K/min \cite{Hartmann2014}. The sudden freezing of the structural degrees of freedom causes a decrease of the electronic bandwidth $W$ with increasing cooling rate, and hence a Mott metal-insulator transition for metallic systems crossing the critical ratio $(W/U)_{c}$ of bandwidth to on-site Coulomb repulsion $U$, see Refs.\ \cite{Hartmann2014,Hartmann2015}. Due to the glassy character of the transition, the effect is persistent below $T_g$ and can be reversibly repeated by melting the frozen configuration upon warming above $T_g$. Both by exploiting the characteristics of slowly-changing relaxation times close to this temperature and by controlling the heating power, the materials can be fine-tuned across the Mott transition \cite{Hartmann2014,Hartmann2015}.\\ 
(ii) The strong EEG vibrations lead to an enhanced scattering contribution at elevated temperatures above $T_g$ \cite{JMueller2009b,Brandenburg2012} and their frozen-in configuration below $T_g$ causes a random lattice potential, which gives rise to an additional contribution to the residual resistivity and affects the superconducting state and transition temperature \cite{Su1998,Su1998a,Yoneyama2004,Powell2004}. 
The influence of quenched disorder on the strongly-correlated $\pi$-carriers, which results in a Mott-Anderson transition, recently has become of increasing theoretical \cite{Shinaoka2009a,Shinaoka2009b} and experimental \cite{Analytis2006,Sasaki2008,Sano2010,Diehl2015} interest.
 
Experimentally, disorder may be realized either by a systematic irradiation with X-rays 
\cite{Sasaki2012} or by 
varying the cooling-rate $q_c = {\rm d}T/{\rm d}t$ through the glass-like transition at $T_g$. 
The latter type of disorder (in the EEG orientations) is intrinsic in nature and can be controlled in a reversible way. Therefore, the cooling-rate dependence may be also utilized as a tool for studying the fundamentals of the influence of randomness on the normal- and superconducting ground state of correlated-electron systems: cooling through $T_g$ with increasing rate systematically suppresses the superconducting $T_c$ of $\kappa$-Br and $\kappa$-CuNCS \cite{Su1998,Su1998a,Stalcup1999}. In Ref.\ \onlinecite{Hartmann2014}, we have described a simple model for \kCl\ and \kBr, where the eclipsed and staggered EEG conformations are described as a double-well potential with an energy difference $2 \Delta E \sim 200$\,K between the E and S state and an activation energy barrier $E_a$ of order $2400 - 3000$\,K. Using this model, we quantitatively estimate the amount of EEG units frozen in the non-equilibrium S orientation, as compared to the majority in the E state, for different cooling rates (see Fig.\,3 in Ref.\ \onlinecite{Hartmann2014}). We find that for moderate and even the rapid cooling rates reported in the literature, the occupation of S conformation is below 5\,\%.\\ 

Besides the well-established experimental phenomenology of the freezing-in of the [C$_2$H$_4$] moieties and despite its significance for the electronic ground state properties, 
the glass-like transition has not been investigated in much detail and its origin in some $\kappa$-(ET)$_2$X salts with polymeric anions -- while no transition has been reported for other (ET)$_2$X compounds -- still remains puzzling. Also, besides a recent NMR investigation on \CuNCS\ \cite{Kuwata2011}, the low-frequency dynamics of the molecular EEG motion itself has not been systematically studied. 
%
The fluctuation spectroscopy data presented in this work, due to the lack of dielectric spectroscopy at elevated temperatures, reveal for the first time characteristic properties specific for orientational glasses like a Vogel-Fulcher-Tammann behavior, which gives insight in the nature of the glass transition in organic charge-transfer salts and allows for a comparison with other glass-forming plastic crystals. {\it Ab initio} studies of the energetics of EEG rotation allow the identification of several key distinctions between those materials displaying glassy freezing.

\section{Methods \label{sec_methods}}
Single crystals of plate- and rod-shaped morphology of $\kappa$-(ET)$_2$X with X = Cu[N(CN)$_2$]Cl, Cu[N(CN)$_2$]Br and Cu(NCS)$_2$, as well as the fully deuterated variant $\kappa$-(D$_8$-ET)$_2$Cu[N(CN)$_2$]Br (denoted as $\kappa$-D$_8$-Br) 
were grown by electrochemical crystallization according to literature methods \cite{Urayama1988,Williams1990,Kini1990,Wang1990,Tokumoto1991}. 


Low-frequency fluctuation spectroscopy measurements have been performed in a five-terminal setup using a standard bridge-circuit ac technique \cite{Scofield1987}. Details of the experiment are described elsewhere \cite{JMueller2011}. Care has been taken that spurious noise sources, in particular contact noise, do not contribute to the results. Resistance $R$ and (resistance) noise power spectral density (PSD) $S_R$ have been measured perpendicular to the conducting layers.
$S_R = S_V/I^2$ is calculated from the measured voltage noise PSD defined by 
\begin{equation}
S_V(f) = 2 \lim\limits_{T \rightarrow \infty} \frac{1}{T} \left| \int\limits_{-T/2}^{T/2} {\rm d}t {\rm e}^{{\rm i}\omega t} \delta V(t) \right|^2, 
\end{equation}
where $I$ and $\delta V(f)$ represent the current through and the fluctuating voltage drop across the sample, resepectively.
For all samples, we have observed excess noise of general $1/f^\alpha$-type, characterizing the intrinsic resistance (conductance) fluctuations. Typical spectra (raw data) which consist of 50 averages and a discussion of reproducibility and error bars are presented in a previous publication \cite{JMueller2009b}. The investigated samples have been continuously cooled down to $4.2\,$K with a slow cooling rate of about $3\,$K/h. For $\kappa$-D$_8$-Br, we also applied a fast cooling rate of $300$\,K/h. Noise spectra have been taken while warming up the sample in discrete steps. The temperature stability during a noise measurement is better than $1\,$mK. The resistance values taken simultaneously with the noise measurements in discrete steps are found to be in very good agreement with the continuous measurements, see Ref. \onlinecite{Brandenburg2012}.

Ab-initio calculations presented in section \ref{Sec4} were performed using ORCA\cite{Neese2012} at the B3LYP/def2-SV(P) level. Starting geometries for a series of ET salts were taken from room temperature crystal structures reported in Ref.~\onlinecite{Hiramatsu2015}. Calculations were performed on isolated ET dimers, with the interaction between the dimer and anion layer accounted for with a OPLS-aa forcefield\cite{Jorgensen1988} implemented in the qmmmopt option. Lennard-Jones parameters were taken from the GROMACS\cite{Pronk2013} set, while the charge distribution of the polymeric anion layer was estimated from Mulliken analysis of B3LYP/def2-SV(P) calculations on small fragments. For each conformation, the coordinates of the EEG of interest were relaxed, along with the two bonded sulfur atoms, which typically display large thermal ellipsoids at room temperature indicative of small disorder. For transition state calculations, ORCA's hessian mode following algorithm was employed.


\section{Experimental Results}
\subsection{Signature of glass-like structural dynamics in resistance noise spectroscopy -- the DDH model}
\begin{figure}[htbp]
\begin{center}
\includegraphics[width=0.8\columnwidth]{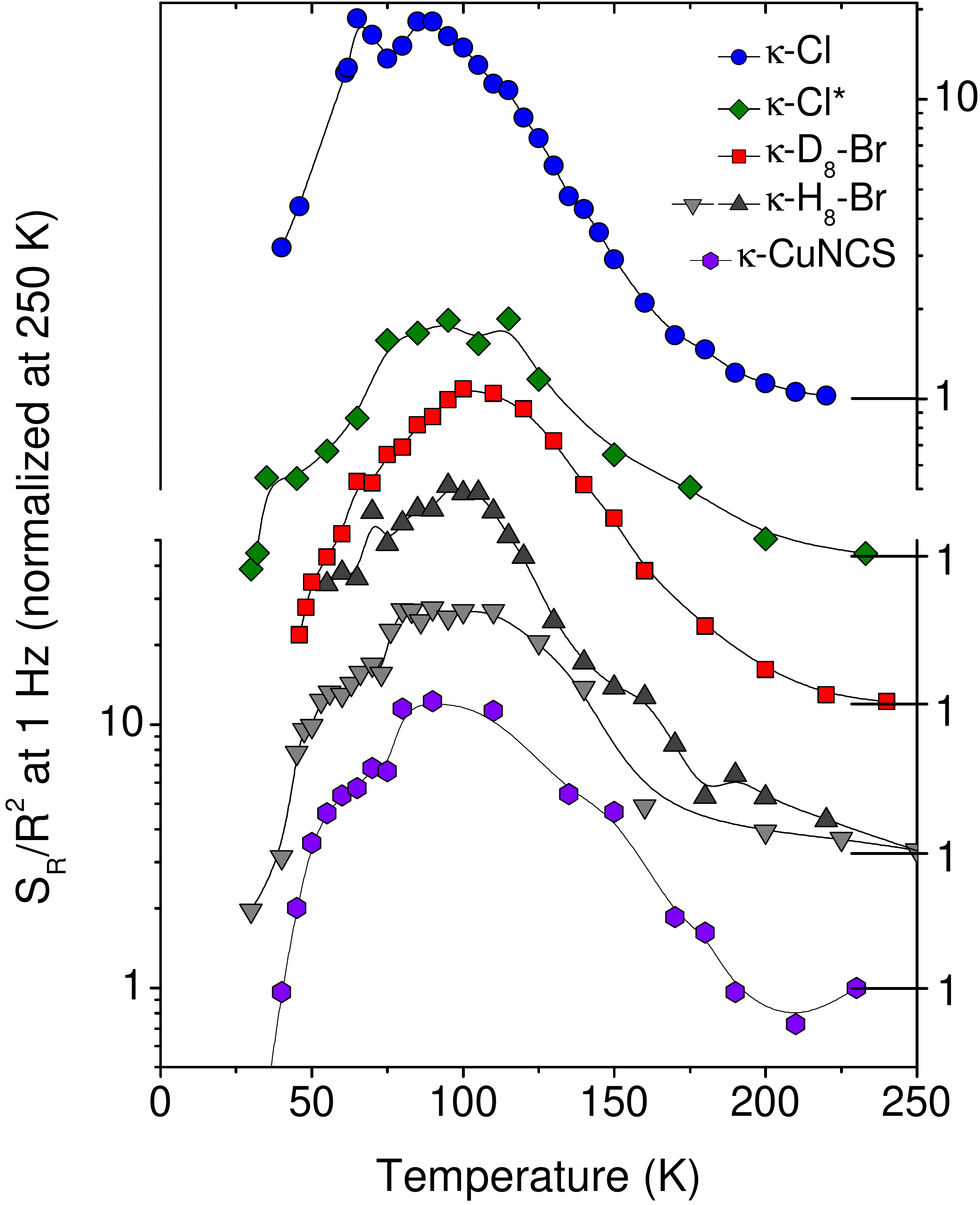}
\caption{\small{(color online) Resistance noise PSD, $S_R/R^2$, taken at $1$\,Hz and normalized to the value at 250\,K, {\it vs.}\ $T$ for five $\kappa$-(ET)$_2$X salts located at different positions in the generalized phase diagram: between insulating $\kappa$-Cl and metallic $\kappa$-H$_8$-Br and $\kappa$-CuNCS, pressurized $\kappa$-Cl$^\ast$ and fully-deuterated $\kappa$-D$_8$-Br are located in the critical region close to the Mott MIT. For $\kappa$-H$_8$-Br data for two different samples are shown. Curves are shifted for clarity. Data have been taken in discrete steps in warm-up measurements after continuously cooling down the samples with slow cooling rates.}}
\label{vergleich_kappa}
\end{center}
\end{figure}
Figure \ref{vergleich_kappa} shows the noise PSD of the resistance fluctuations, $S_R/R^2$, taken at 1\,Hz and normalized to the value at $T = 250$\,K, for five compounds of the $\kappa$-(ET)$_2$X family located at different positions in the generalized phase diagram, 
and thus having different electronic ground states. $\kappa$-Cl is insulating under ambient conditions and is shifted to the critical region close to the Mott MIT by application of physical pressure ($\kappa$-Cl$^\ast$, same sample as in Ref.\ \onlinecite{JMueller2009b}). Likewise, chemically-induced pressure in fully-deuterated $\kappa$-D$_8$-Br (same sample as in Ref.\ \onlinecite{Brandenburg2012}) drives the ambient-pressure superconductor $\kappa$-H$_8$-Br close to the Mott transition. $\kappa$-H$_8$-Br and $\kappa$-CuNCS are metallic.
Strikingly, in spite of the very different electronic ground-state properties determined by the electronic correlation strength $W/U$, the behavior at elevated temperatures of the different compounds is very similar, namely governed by a pronounced maximum in the noise power centered around $T \sim 100$\,K. Besides the broad maximum, the data sets display an anomalous feature at the thermodynamic glass-transition temperature $T_g$, {\it i.e.}\ a peak or small shoulder in $S_R/R^2(T)$. It is natural to conclude that the overall temperature dependence is due to the structural similarities of the different compounds. 

\begin{figure}[htbp]
\begin{center}
\includegraphics[width=\columnwidth]{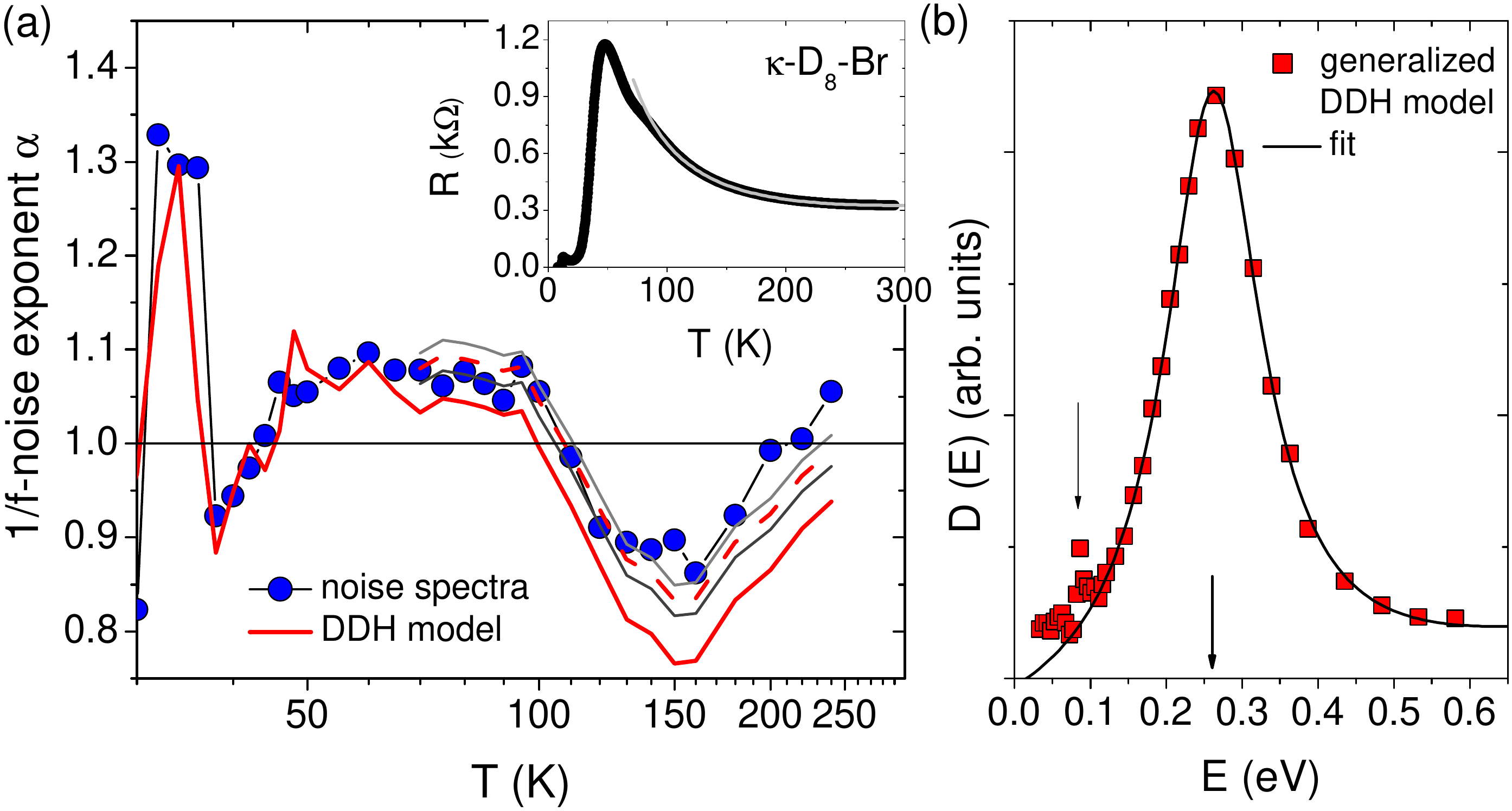}
\caption{\small{(color online) (a) Temperature dependence of the $1/f^\alpha$-noise frequency exponent, $\alpha(T)$, for $\kappa$-D$_8$-Br. Blue dots represent the data, the colored curves show the prediction $\alpha_{\rm calc}$ after Eq.\,(\ref{eq2}) with $g(T) = {\rm const.}$ (solid red line) and for the generalized DDH model with $g(T) = a T^{b}$ for $T \geq T_{g}$ with $b =  - 3/2$ (dashed red line), $b = - 1$ (dark grey) and $b = - 2$ (light grey). Inset shows the temperature dependence of the resistance, which above $T_g$ has been fitted to $R(T) = R_0 + R_1T + R_{3/2}T^{-3/2}$. (b) Distribution of activation energies $D(E,f= 1\,{\rm Hz})$ in the generalized DDH model, {\it i.e.} using $g(T) = aT^{-3/2}$ for $T \geq T_g$ in Eq.\,(\ref{eq3}). Line is a fit to a Lorentz peak function and a linear term and serves as guide to the eye (a Gaussian fit works equally well in the region around the maximum, see also Fig.\,\ref{frequenzen}). Thin arrow points to enhanced fluctuations induced by electronic correlations \cite{Brandenburg2012}. Thick arrow marks the Energie $E_{\rm max} \approx 260$\,meV.}}
\label{DDH_Daten}
\end{center}
\end{figure}

Besides the magnitude of the $1/f$-type noise shown in Fig.~\ref{vergleich_kappa}, it is interesting to analyze its frequency dependence. 
In a previous study \cite{Brandenburg2012} we have demonstrated that the temperature dependence of the observed excess noise can be explained by the phenomenological Dutta-Dimon-Horn (DDH) model \cite{Dutta1979}, which assumes that the $1/f^\alpha$-type noise is caused by many independent `fluctuators' -- each contributing a Lorentzian PSD -- acting in concert. 
These fluctuators may be identified with thermally-activated two-level processes characterized by a relaxation time $\tau=\tau_0 \exp{(E/k_{\rm B}T)}$, where $\tau_0\approx (10^{-12}-10^{-14})$\,s is an attempt time in the order of the inverse phonon frequencies, $E$ the activation energy of the process and $k_{\rm B}$ Boltzmann's constant. 
%
The voltage autocorrelation function for such a two-level system, 
\begin{equation}
\mathcal{A}_V(t) = \langle \delta V(t + \tau) \cdot \delta V(t) \rangle = \overline{(\delta V)^2}\exp{(-|t|/\tau)}, 
\end{equation}
is purely exponential, while the PSD, according to the Wiener-Khintchine theorem, is a Lorentzian function of frequency \cite{Machlup1954,Kogan1996}:
\begin{equation}
S_V(f) = 4 \int_0^\infty \mathcal{A}_V(t) \cos{(2 \pi f t)} {\rm d}t =   \overline{(\delta V)^2} \frac{4 \tau}{1+4\pi^2 f^2 \tau^2}.
\label{two-level}
\end{equation}
In a particular 'noise window' the action of an individual fluctuator may be enhanced over the $1/f$ background, such that the kinetics and characteristic energy of this process can be determined \cite{JMueller2009a}.\\
The DDH model assumes that the {\it a priori} unspecified fluctuators linearly couple to the measured resistance fluctuations. The main ingredient of the model is a distribution of activation energies \(D(E)\), which governs both the strong temperature dependence of the magnitude of the noise, $S_R/R^2(f,T)$, and any deviations from perfect ($\alpha = 1$) $1/f$ behavior, {\it i.e.} the temperature dependence of the $1/f^\alpha$-noise frequency exponent, $\alpha(T)$. In this model, it is
\begin{equation}
\frac{S_R}{R^2}(f,T) \propto \int g(T) \frac{}{} \frac{\tau_0 \exp{(E/k_{\rm B}T)}}{1+\tau_0^2 \exp{(2E/k_{\rm B}T)}4\pi^2f^2} D(E) \mathrm{d}E,
\label{eq1}
\end{equation}
and
\begin{equation}
\alpha_{\rm calc}(f,T) = 1 - \frac{1}{\ln 2 \pi f \tau_0} \biggl(\frac{\partial \ln{\frac{S_R}{R^2}(f,T)}}{\partial \ln T} - \frac{\partial \ln g(T)}{\partial \ln T} - 1\biggr).
\label{eq2}
\end{equation}
Assuming that $D(E)$ is a slowly varying function compared to thermal energies $k_{\rm B} T$, it follows that
\begin{equation}
D(E) \propto \frac{2 \pi f}{k_{\rm B} T}\frac{1}{g(T)}\frac{S_R}{R^2}(f,T).
\label{eq3}
\end{equation}
Due to the large logarithmic factor in \(E = - k_{\rm B} T \ln{(2 \pi f \tau_0)}\), ordinary activation energies in solids can be accessed \cite{Kogan1996}. The strength of this approach is that -- if \(D(E)\) in Eq.\,(\ref{eq1}) does not explicitly depend on temperature -- the assumptions of the DDH model can be checked for consistency by comparing the measured frequency exponent 
\begin{equation}
\alpha_{\rm meas}(T) = - \frac{\partial \ln{\frac{S_R}{R^2}(f,T)}}{\partial \ln{f}} 
\label{eq2a}
\end{equation}
with the predictions $\alpha_{\rm calc}(T)$ of the model from Eq.\,(\ref{eq2}) \cite{Dutta1979}. We discuss here a generalization of the DDH model \cite{Black1983,Fleetwood1984,Raquet1999}, which accounts for possible deviations of the predicted (calculated) frequency exponents from the measured data: a function $g(T)$ may account for an explicit temperature dependence of the distribution of activation energies. The physical meaning is that the coupling constant between the random processes and the resistance, and hence the total noise magnitude $\overline{(\delta R)^2} = \int_0^\infty S_R(f){\rm d}f$, is {\it not} independent of temperature. In the simplest case of a constant offset (vertical shift) of $\alpha_{\rm calc}(T)$ as compared to $\alpha_{\rm meas}(T)$, $g(T)$ satisfies a simple power law $g(T) = aT^b$, where the exponent $b$ is given by the vertical shift and describes the temperature dependence of the number and/or coupling strength of the fluctuators. Fig.\,\ref{DDH_Daten}\,(a) shows, exemplarily for $\kappa$-D$_8$-Br, an excellent agreement of the measured data with the DDH model \cite{Brandenburg2012}, when for higher temperatures $T \geq T_g$ the almost constant deviation between the measured and calculated frequency exponent is accounted for by a power law $g(T) = aT^b$, where $b = -3/2$ gives a reasonably good description. Hints for a physical interpretation of this power-law bevaviour comes from the temperature dependence of the resistance shown in the inset, where above $T_g$ a scattering contribution $\propto T^{-3/2}$ can be identified, 
suggesting that the responsible mechanism may be linked to the function $g(T)$ via the number and/or strength of the fluctuators, similar as has been argued for the contribution of spin fluctuations in certain manganites \cite{Raquet1999}. In our case, the fluctuators at elevated temperatures can be associated with structural excitations involving the EEG, the coupling of which to the resistivity \cite{Kuwata2011} varies as $T^{-3/2}$, see Ref.\ \onlinecite{comment2}.

The non-monotonic temperature dependence $\alpha(T)$ shown in Fig.\,\ref{DDH_Daten}\,(a) is related to the energetic fingerprints of the fluctuating entities, since for $\alpha$ greater and smaller than 1, $\partial D(E)/\partial E > 0$ and $\partial D(E)/\partial E < 0$, respectively. The good agreement of $\alpha_{\rm meas}(T)$ with the generalized DDH model allows for extracting \(D(E)\) from the temperature dependence of the magnitude of the resistance noise PSD by using Eq.\,(\ref{eq3}). Figure\,\ref{DDH_Daten}(b) shows $D(E,f = 1\,{\rm Hz})$, calculated with $g = aT^{-3/2}$ for $T \geq T_g$. The features below $E \sim 0.1$\,eV (see arrow) corresponding to the sharp peak in $\alpha(T)$, which is also accounted for by the DDH model, are due to a sudden slowing down of the charge carrier dynamics in the vicinity of the critical endpoint of the Mott MIT as discussed in Refs.\ \cite{Brandenburg2012,Hartmann2015}. Besides these electronic correlation effects at lower energies, $D(E,f=1\,{\rm Hz}$ can be reasonably well described by a small linear term and a peak function (a Lorentzian function is shown in Fig.\,\ref{DDH_Daten}(b); a Gaussian peak function works equally well around the maximum, see Fig.\,\ref{frequenzen} below) with the maximum at $E_{\rm max} \approx 262$\,meV ($\approx 3040$\,K) and full-width-at-half-maximum of $\sim 160$\,meV ($\sim 1860$\,K) \cite{comment3}. As we have discussed in our previous studies, $E_{\rm max}$ corresponds to the activation energy $E_a$ of the orientational degrees of freedom of the ET molecules' terminal ethylene groups, as determined e.g.\ by NMR, specific heat and thermal expansion \cite{Miyagawa1995,Wzietek1996,Akutsu2000,Mueller2002}. Thus, in this case the relevant fluctuators can be identified {\it a posteriori} as the ET molecules' EEG  \cite{JMueller2009b}. Thus, the strong resistance fluctuations at around 100\,K shown in Fig.\,\ref{vergleich_kappa} originate in the strong coupling of the electronic degrees of freedom to the glass-like structural excitations, see also \cite{Kuwata2011}. The corresponding change in the frequency exponent $\alpha(T)$ upon cooling from $\alpha < 1$ to $\alpha > 1$ corresponds to a strong shift of spectral weight to lower frequencies. This is related to the expected slowing down of molecular dynamics when approaching the glass transition and consistent with our interpretation discussed below, namely that the observed charge carrier dynamics is caused by the slow, so-called $\alpha$-process of the glass-forming EEG rotations embedded in the anion structure forming a 'cage' environment through short C$-$H\,$\cdots$\,anion contacts \cite{Geiser1991}. In dielectric spectroscopy, the $\alpha$-relaxation peak of the frequency-dependent dielectric loss -- which may be connected to $S(f)$ by a fluctuation-dissipation relation -- shifts with decreasing temperature to lower frequencies in a similar way. 

\subsection{Analysis of the dynamic transport properties}
Next we show that noise measurements allow for studying characteristic properties of the glassy system as, {\it e.g.}, a possible non-Arrhenius behavior, which
so far has not been investigated for the present materials. The frequency range of the $1/f$-type fluctuations for the present materials is naturally limited to about $1\,{\rm mHz} - 1\,{\rm kHz}$. In this range, the relaxation is dominated by the $\alpha$-relaxation, which is the slowest dynamic process in glass formers. In contrast to other, faster processes of different origin (as, {\it e.g.}, the high-frequency $\beta$-process, 
which can be visualized as a 'rattling' movement of a particle in the transient 'cage' formed by its surrounding), the $\alpha$-relaxation is associated with the forming and decay of the cage, involving the cooperative movement of many particles \cite{Lunkenheimer2000}, in our case the ET molecules' EEG and their anion environment.
The increase of the $\alpha$-relaxation time during cooling and deviations from a thermally-activated behavior in glass-forming materials is often parameterized by
an empirical law employed by Vogel, Fulcher and Tammann 
\cite{VFT1926,Angell_1985}:
\begin{equation}
\tau = \tau_0 \exp{\left[\frac{D \cdot T_{\rm VFT}}{T - T_{\rm VFT}}\right]},
\label{VFT}
\end{equation}
where $\tau_0$ is a prefactor, $D$ is the so-called strength parameter and $T_{\rm VFT}$ is the Vogel-Fulcher-Tammann temperature. The divergence of the relaxation time at $T_{\rm VFT}$ cannot be observed in a real experiment because the system falls out of equilibrium when the relaxation time becomes larger than the time scale of the experiment. The model suggests a 'hidden' phase transition below  the glass transition at $T_g$, which, however, is controversially discussed. 
The strength parameter $D$ can be used to classify the glass forming materials in terms of the deviations of $\tau(T)$ from a thermally activated behavior, for which $T_{\rm VFT} = 0$ \cite{Angell1995,Boehmer1993,Lunkenheimer2000}. So-called 'fragile' glass formers reveal small values of $D$ (typically $D < 10$) with large deviations from an Arrhenius behavior, whereas 'strong' glass formers ($D > 10$) follow an Arrhenius law more closely. This classification scheme has proven very useful since many properties of glass-forming materials are correlated with their strength or fragility. Typical fragile glass formers are substances with nondirectional interatomic/intermolecular bonds, {\it e.g.}\ molten salts. Strong systems, {\it i.e.}\ those which show a strong 'resistance' against structural degradation
when heated through their supercooled regime, are found among the non-hydrogen bonded network melts \cite{Boehmer1993}.
Another convenient measure for the strength of the fragility can be obtained by focusing on the long relaxation time end of the strong-fragile pattern \cite{Boehmer1993,Wang2006}. The (dimensionless) slope at $T_{g}$ is given by
\begin{equation}
m = \frac{{\rm d}\log{\tau}}{{\rm d}(T_{\rm g}/T)}\Big|_{T=T_{g}}
\label{slope}
\end{equation}
and is related to the strength parameter $D$ by
\begin{eqnarray}
m & = & (D/\ln{10})\cdot(T_{\rm VFT}/T_{\rm g})\cdot\left(1-T_{\rm VFT}/T_{\rm g}\right)^{-2} \\
    & = & m_{\rm min}+m_{\rm min}^2{\rm ln}10/D \nonumber 
\end{eqnarray}
where $m_{\rm min}\approx 16$ is found to be the lower limit of fragility \cite{Boehmer1993}. For a large class of glass-forming liquids, the upper bound for the fragility index is $m_{\rm max} = 170$ \cite{Wang2006}.

\begin{figure}[htbp]
\begin{center}
\includegraphics[width=\columnwidth]{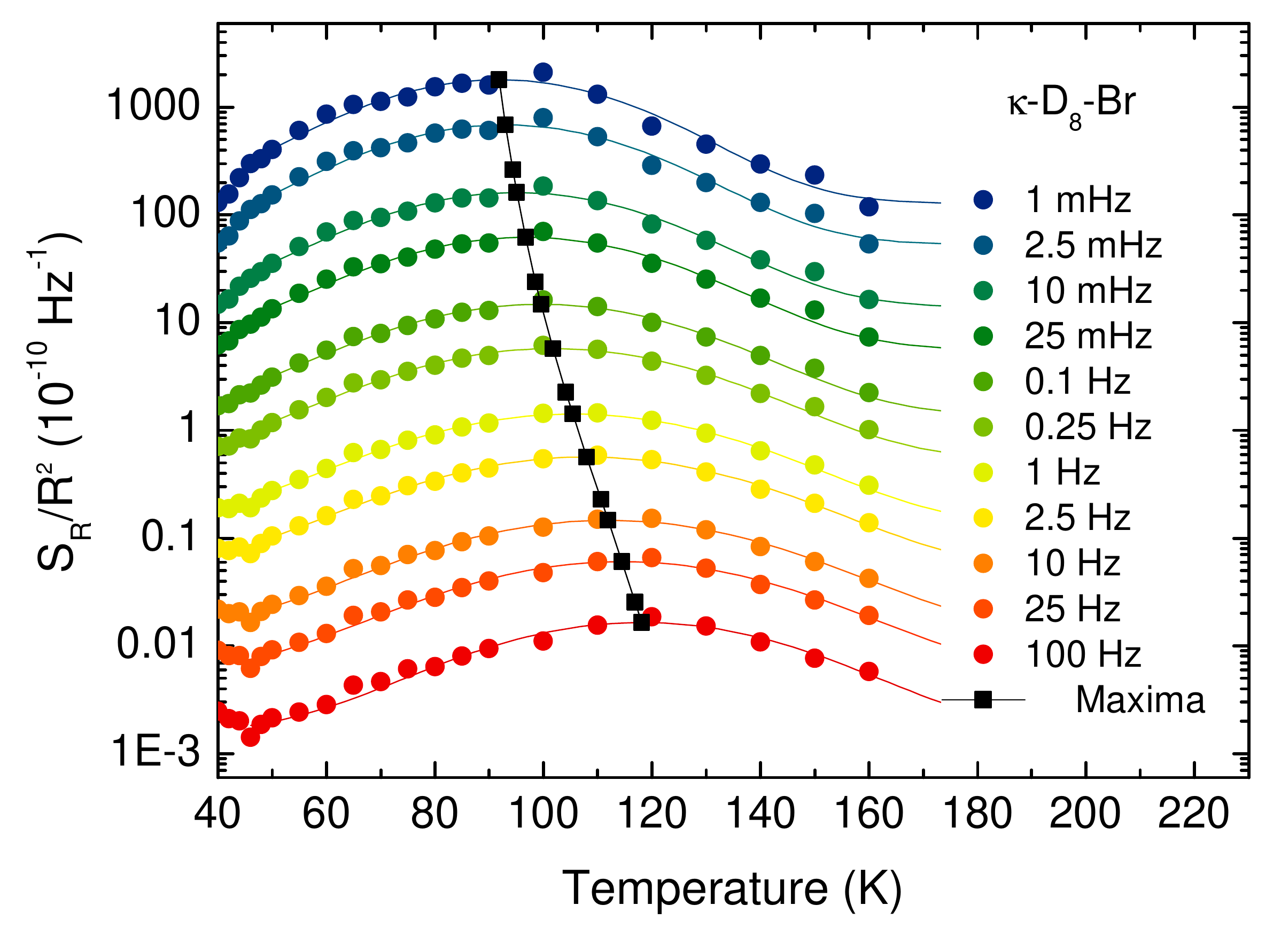}
\caption{\small{(color online) Data (circles) of the normalized resistance noise PSD recalculated from the $1/f^\alpha$ frequency exponent $\alpha(T)$ shown in Fig.\,\ref{DDH_Daten}(a) and Gaussian fits (lines) {\it vs.}\ temperature for various frequency of $\kappa$-D$_8$-Br. (The distribution of activation energies $D(E)$ in the DDH model shown in Fig\,\ref{DDH_Daten}(b) are derived from the data for $f = 1$\,Hz.) The black squares connecting the maxima of the fits represent the temperature dependence of the relaxation time characteristic for the EEG structural excitations.}}
\label{frequenzen}
\end{center}
\end{figure}
\begin{figure}[htbp]
\begin{center}
\includegraphics[width=\columnwidth]{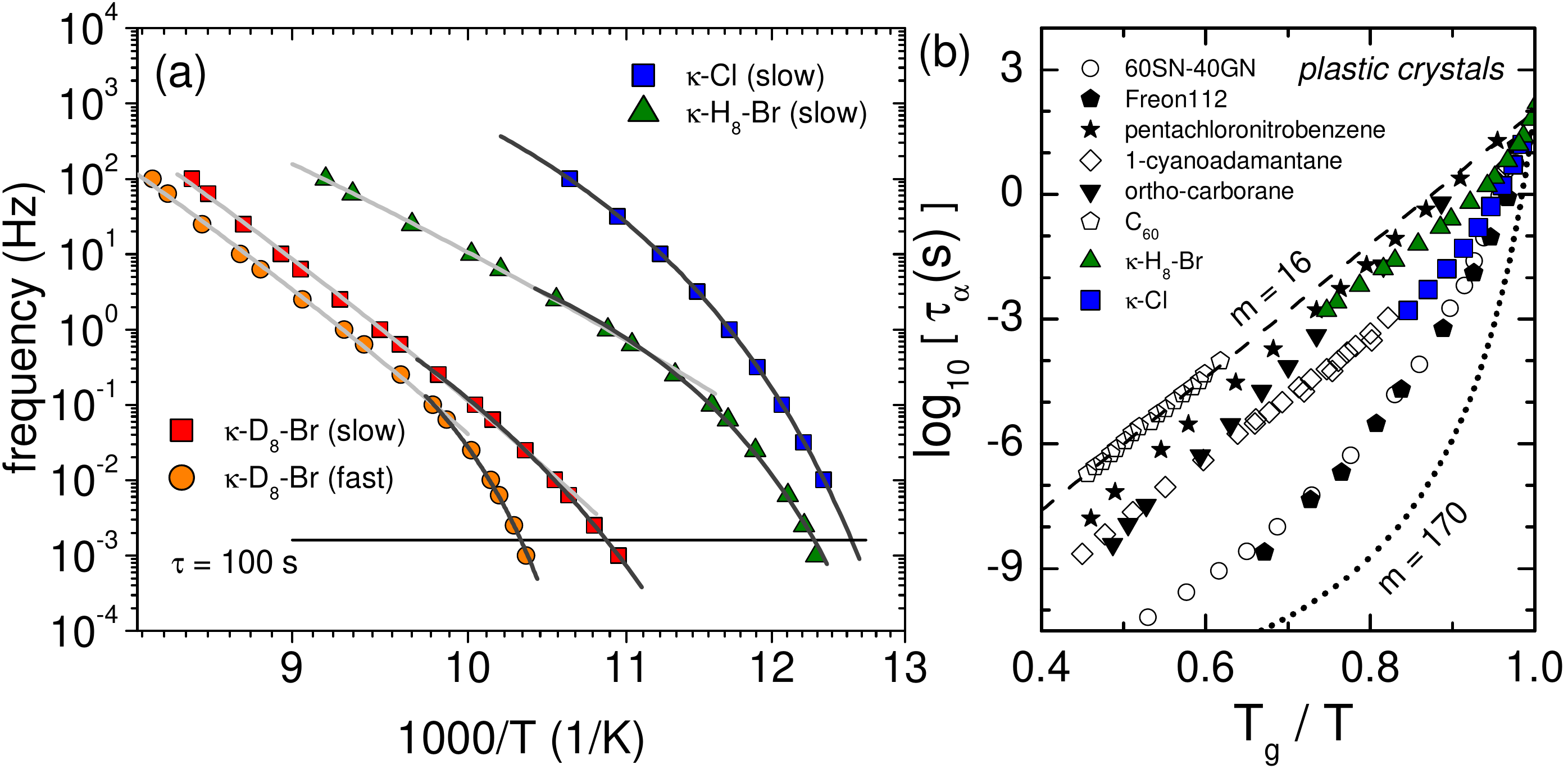} 
\caption{\small{(color online) (a) Arrhenius plot of the noise peak frequency as shown exemplarily in Fig.\,\ref{frequenzen} for selected systems of $\kappa$-(ET)${}_{2}$X (see legend). Light grey lines are fits of thermally-activated (Arrhenius) behavior to the high-frequency data, whereas dark grey lines are VFT fits after Eq.\,(\ref{VFT}) to the low-frequency data. The horizontal line equals $\tau = 100$\,s. (b) The data for $\kappa$-H$_8$-Br and \kCl\ in comparison to various other orientational glasses (taken from Ref.\ \onlinecite{Brand2002}).}}
\label{Vogel-Fulcher}
\end{center}
\end{figure}

Here, we employ the Vogel-Fulcher-Tammann expression, Eq.\,(\ref{VFT}), in order to describe the temperature dependence of the peak in the $1/f^\alpha$-noise PSD caused by the glass-like structural excitations at a certain frequency. The frequency dependence is essentially given by the noise exponent $\alpha(T)$ shown in Fig.\,\ref{DDH_Daten}(a) for $\kappa$-D$_8$-Br. The recalculated noise amplitude $S_R/R^2(T)$ for different frequencies is shown in Fig.\,\ref{frequenzen}. The black squares that connect the maxima of the Gaussian fits thus represent the temperature dependence of the relaxation time characteristic for the EEG\,$\cdots$\,anion structural excitations. In Fig.\,\ref{Vogel-Fulcher}(a) we plot these maxima at different frequencies in an Arrhenius plot for different systems. Strikingly, the curves follow a thermally-activated behavior for high frequencies, in agreement with earlier measurements of thermodynamic properties \cite{Mueller2002}. For lower frequencies, however, a clear deviation from an Arrhenius behavior is observed and the curvature can be described well by the Vogel-Fulcher-Tammann law, Eq.\,(\ref{VFT}). The corresponding slowing down of the motion of the structural units being significantly stronger than predicted by the Arrhenius law is a typical characteristic of a glassy system \cite{Bauer2013}. As we will discuss below, a possible explanation for this behavior is a growing number of correlated molecules moving cooperatively which leads to an increase of the apparent activation energy for decreasing temperatures.\\
We note that the definition of the glass transition temperature is not unambiguous since it depends on the cooling rate. A possible procedure for defining $T_g$ from thermodynamic measurements is described above.
In spectroscopy measurements, $T_g$ is often defined as the temperature where the relaxation time becomes $\tau = 100$\,s, see Refs.\ \cite{Jaeckle1986,Stickel1995,Lunkenheimer2000,Debenedetti2001}. 
Equivalently, in order to have a well-defined transition temperature for different materials, $T_g$ is determined as the temperature, where the system falls out of equilibrium for $q = 10$\,K/min. Although we have employed cooling rates different than $q = 10$\,K/min, the corresponding frequency $f = 1/(2\pi\tau) \approx 1.6$ mHz is shown as a horizontal line in Fig.\,\ref{Vogel-Fulcher}(a).
As expected, the Vogel-Fulcher-Tammann temperature $T_{\rm VFT}$ is lower than the glass transition temperature $T_{\rm g}$ with the VFT-fits in Fig.\,\ref{Vogel-Fulcher}(a) for the various slowly-cooled systems yielding $T_{\rm VFT} = 65$\,K  for \kCl, 
$T_{\rm VFT} = 64$\,K and 85\,K for slowly ($q = 3$\,K/h) and fast ($q = 300$\,K/h) cooled $\kappa$-D$_8$-Br, respectively, 
and 68\,K and 64\,K for two different samples of $\kappa$-H$_8$-Br. 
The cooling-rate dependence of the VFT behavior, observed for $\kappa$-D$_8$-Br, is unusual since for classical glass-forming systems the curves for different $q$ should coincide for $T > T_g$. Rather than being a purely dynamical effect, we suggest that for the present glass-like materials different cooling rates produce different glass-forming molecular environments (see discussion below) resulting in different values of $T_g$.\\  
Orientational glasses usually are 'strong' glassformers, in the sense that they exhibit approximately an Arrhenius behavior in a plot of the viscosity $\eta$ of the liquid $\ln{\eta}$ {\it vs.}\ $T_g/T$, indicative of a temperature-independent activation energy \cite{Hoechli1990}. 
Fragile liquids on the other hand exhibit super-Arrhenius behavior, {\it i.e.}\ their effective activation energy increases as temperature decreases.
From our analysis of the noise measurements, we find a wide range of the strength parameter with $D = 5$ for \kCl, $D = 8.6$ and $D = 1.8$ for slow- and fast-cooled $\kappa$-D$_8$-Br, respectively and $D = 2.8$ and $5.2$ for two different samples of $\kappa$-H$_8$-Br (only one shown in Fig.\,\ref{Vogel-Fulcher}). Apart from obvious sample-to-sample dependencies and errors related to the limited number of data points, values below $D = 10$, consistently observed for a total of 6 samples, reveal that the present charge-transfer salts are rather fragile glasses, indicating significant cooperativity between EEG degrees of freedom. Accordingly, the values for the slope are determined to be in the range of $m \sim 100 - 200$ for the different samples, {\it i.e.}\ in each case much larger than the lower limit of fragility corresponding to $m = 16$ (Ref.\ \onlinecite{Boehmer1993}). 
For comparison, in Fig.\,\ref{Vogel-Fulcher}(b) we plot the data for $\kappa$-H$_8$-Br and $\kappa$-Cl in the classification scheme with other plastic crystals or orientational glasses \cite{Brand2002}. 

Besides the VFT behavior, a typical characteristic of glass formers is non-exponential relaxation, which may be caused by a distribution of relaxation times and usually is seen in an asymmetric broadening of the peaks in the dielectric loss \cite{Lunkenheimer2000}. In general, the fluctuation-dissipation theorem relates the noise PSD of a quantity $x$ to the imaginary part of the complex susceptibility $A^{\prime \prime}(\omega)$, where the real function $A(t)$ is the response function of the system \cite{Kogan1996}:  
\begin{equation}
S_x(f) = 2k_{\rm B}T \frac{A^{\prime \prime}(\omega)}{\omega}.
\label{FDT}
\end{equation}
Despite our limited frequency range, transforming the data of Fig.\,\ref{frequenzen} according to Eq.\,(\ref{FDT}) reveals -- as expected for a glass-forming system -- asymmetric and considerably broader peaks than 1.14 decades half width given by the Debye equation \cite{Lunkenheimer2000}.

\section{Discussion \label{Sec4}}
\begin{figure}[t]
\begin{center}
\includegraphics[width=\columnwidth]{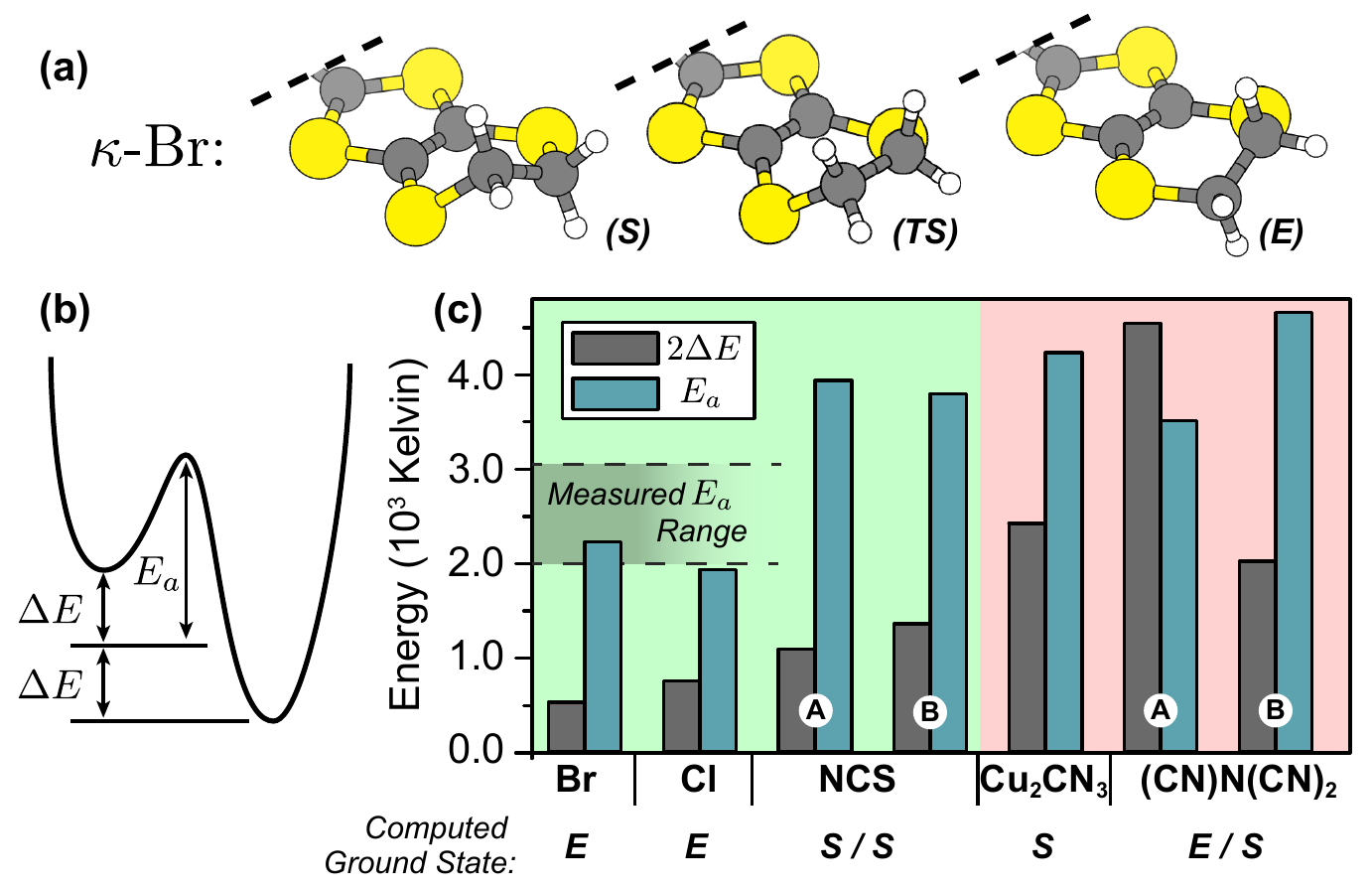}
\caption{\label{Fig7}\small{(color online) 
(a) Optimized geometries of the terminal EEG for $\kappa$-Br in the staggered (S), eclipsed (E), and intermediate transition state (TS) conformation. 
(b) Schematic two-level potential energy surface defining $E_a$ and $\Delta E$. (c) Computed ground state, $E_a$ and $2 \Delta E$ values for selected (ET)$_2$X salts. For those salts with nonequivalent EEGs, the specific sublattice is indicated by an "A" or "B". Only those salts on the left of the figure, indicated by a green background, display glass transitions.}}
\label{anion_structure}
\end{center}
\end{figure}

After having described the phenomenology of the glass-like EEG ordering in resistance noise spectroscopy, we now discuss the question of the occurrence and origin of the effect in $\kappa$-(ET)$_2$X. While enhanced electron scattering associated with thermally-excited structural fluctuations in the EEG orientational degrees of freedom may be a common feature of many (ET)$_2$X salts, a pronounced glass transition has been reported in the literature for the present $\kappa$-Cl, $\kappa$-Br, and $\kappa$-CuNCS systems. 
%
%
In all three materials (and deuterated variants), the planar anion layer consists of one-dimensional Cu(I) coordination polymers (repeat unit X = Cu$^+$[Y$^-$][Z$^-$]) with bridging ligands (Y) completing the backbone, and short side-chains formed by terminal ligands (Z) coordinated to each Cu(I). For example, the isostructural $\kappa$-Cl and $\kappa$-Br crystallize in the space group $Pnma$, with zigzag anion chains consisting of Y = dicyanamide, [N(CN)$_2$]$^-$, and Z = Cl$^-$ or Br$^-$, while $\kappa$-CuNCS (for which Y = Z = thiocyanate, (SCN)$^{-}$) crystallizes in the $P2_1$ space group. Other salts in this latter space group, such as $\kappa$-(ET)$_2$Cu(CN)[N(CN)$_2$], i.e. Y = (CN)$^-$, Z = [N(CN)$_2$]$^-$, apparently show no evidence for structural disorder in the EEGs, let alone a glass transition. 
Curiously, thermal disorder of the EEGs has been found at room temperature in the $P2_1/c$ quantum spin liquid material $\kappa$-Cu$_2$CN$_3$, although thermodynamic measurements suggest a smooth ordering into the staggered conformation at low temperatures without a pronounced glass transition.\cite{deSouza2015} In this latter material, the anion layer forms a more rigid 2D coordination polymer in which (CN)$^-$ anions bridge adjacent -CuCN- chains. Given the structural similarities of the ET layers in each material, the potential for glass formation must originate in the specific nature of the coupling between the anion layer and EEG degrees of freedom. Indeed it has been suggested that such coupling is necessary for the strong cooperative dynamics between EEGs indicated by the here observed fragile glass response (D $<$ 10), as the anion layer is necessary to mediate EEG-EEG interactions over long range\cite{Akutsu2000,Saito1999,Wolter2007}. In order to explore the qualitative effects of the EEG-anion coupling between these materials, the energetics of the E and S conformations has been investigated at the B3LYP/def2-SV(P) level using ORCA, and treating the anion-cation interaction in the static OPLS-aa approximation, as described in Sec. \ref{sec_methods}. The results are summarized in Fig. \ref{Fig7}(c) in terms of the computed ground states (E or S), energy differences $2\Delta E = |E_\text{S}-E_\text{E}|$, and activation energies estimated from $(E_a = E_{\text{TS}} - (E_\text{S}+E_\text{E})/2)$, where `TS' refers to the transition state. In all cases, the computed ground state agrees qualitatively with the observed conformational preferences.

\begin{figure}[t]
\begin{center}
\includegraphics[width=0.75\columnwidth]{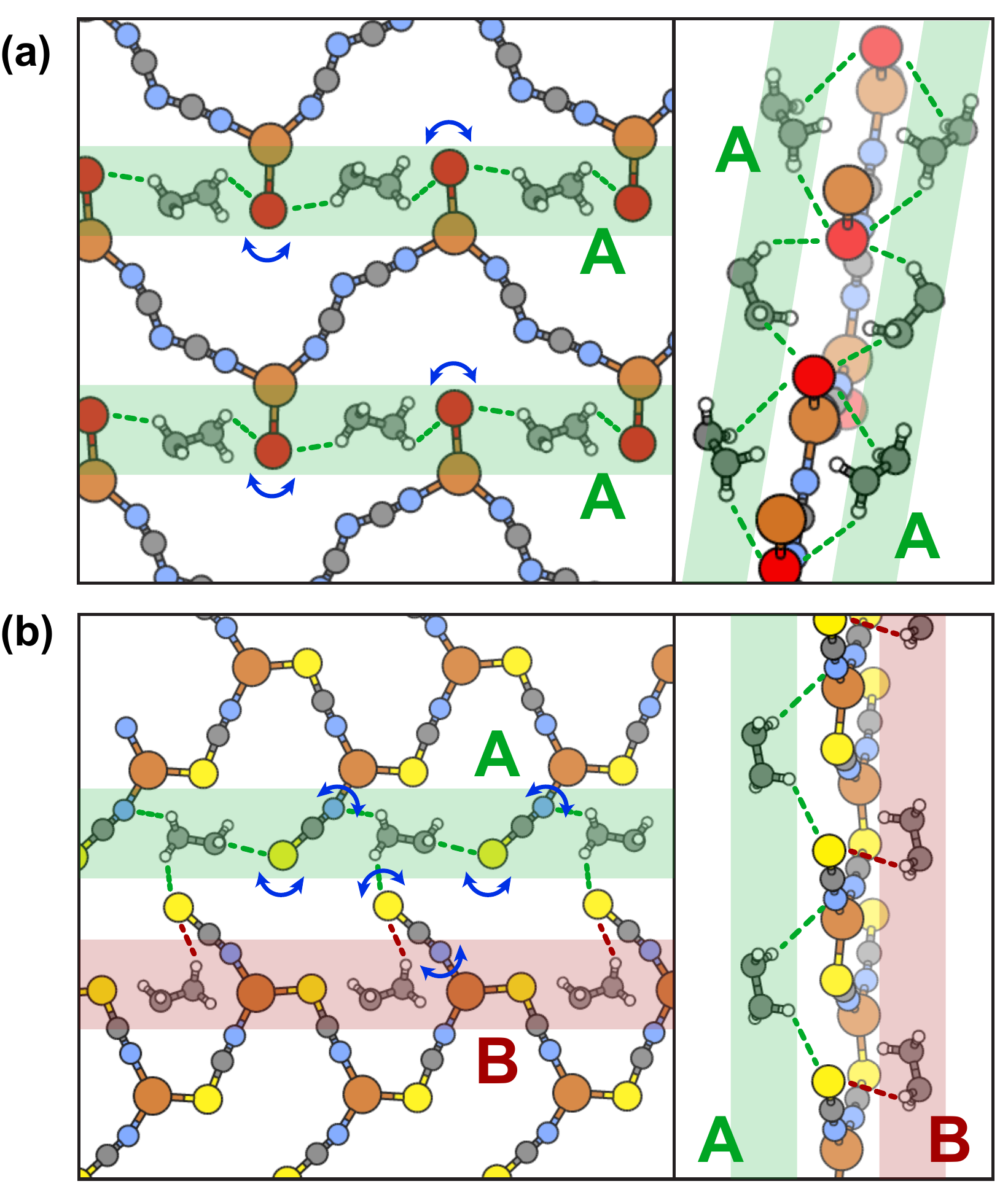}
\caption{\label{Fig8}\small{(color online) Preferred conformations of the EEGs for (a) $\kappa$-Br and (b) $\kappa$-CuNCS in relation to the nearby anion layer. Dashed lines indicate close EEG $\cdot\cdot\cdot$ Z contacts, while blue arrows indicate soft vibrational degrees of freedom of the terminal Z ligands suggested to couple to the EEG rotation. In $\kappa$-Br, all EEG are crystallographically equivalent, while $\kappa$-CuNCS consists of two unique A- and B-sublattices.}}
\label{anion_structure}
\end{center}
\end{figure}

For the isostructural $\kappa$-Cl and $\kappa$-Br, the EEGs exhibiting disorder are all crystallographically equivalent, and were computationally found to prefer the E conformation by $2\Delta E/k_{\rm B} \sim$ 750 K and 520 K per ET molecule for $\kappa$-Cl and $\kappa$-Br, respectively. Despite qualitative agreement with experiment, these calculated energy differences are somewhat larger than the value of $2\Delta E/k_{\rm B} \sim$ 210 K estimated from the temperature dependence of the relative E/S occupancy of $\kappa$-Br for fixed cooling rate\cite{Hartmann2014}. The computed $E_a$ values of 1930 K (Cl) and 2200 K (Br) agree well with the experimental range of $E_a/k_{\rm B} =$ 2000$-$3100 K as determined e.g.\ by NMR, specific heat, thermal expansion, and resistivity measurements of structural relaxation kinetics on these materials\cite{Miyagawa1995,Wzietek1996,Akutsu2000,Mueller2002}. For both salts, more than half of the computed  energy differences $2\Delta E$ result from strong van der Waals coupling between the EEG and anion layers, which is dominated by close Z $\cdot\cdot\cdot$ H (or D) contacts between each EEG and two terminal ligands in adjacent anionic chains (Fig. \ref{Fig8}(a)). In turn, each Z = Br$^-$ or Cl$^-$ interacts with four EEGs, two in each adjacent ET layer. These terminal ligands exhibit significant thermal motion at room temperature, particularly parallel to the chain direction, as evidenced by a large root mean square displacement of $\sqrt{U_{aa}}$ = 0.29 \AA \ for $\kappa$-Br and 0.25 \AA \ for $\kappa$-Cl. The coupling of this motion with the EEG rotation in nearby ET molecules provides a possible mechanism for the cooperative behavior implied by the observed deviations from the Arrhenius law in Fig.\,\ref{Vogel-Fulcher}. 
In this way, it is the collective motion of EEG and anions that freezes out at $T_g$ instead of being simply a property of the vibrational degrees of freedom of the EEG within each individual, separated ET layer. Recently, it has been shown that for glass-forming liquids an increase of the number of correlated molecules causes the noncanonical (non-Arrhenius) slowing down of the molecular motion upon approaching $T_g$ \cite{Bauer2013}, which we also observe for the present materials. Furthermore, the activation energy was found to be approximately proportional to the number of cooperatively moving structural units. This may also be the case for the EEG motion and may be related to the change of the lattice constant with temperature. Spatially-resolved measurements, as {\it e.g.}\ scanning tunneling microscopy, would be highly desirable in order to investigate if there is an inhomogeneous distribution of the EEG orientations with ordered domains as suggested in Ref.\ \onlinecite{Yoneyama2004}.

The computational results for $\kappa$-Cl and $\kappa$-Br may be contrasted with those of the 'spin liquid' compound $\kappa$-Cu$_2$CN$_3$. In the latter material, all disordered EEGs are also crystallographically equivalent, due to the presence of inversion centres in the $P2_1/c$ space group. However, appearing on such an inversion centre in the anion layer is a (CN)$^-$ ion whose orientation (-CN- or -NC-) is statistically disordered. For this reason, the EEG-anion coupling was computed from the average of both cyanide orientations. In agreement with experimental findings, a preference for the S conformation was found, albeit with a much larger $2\Delta E/k_{\rm B} \sim$ 2410 K, suggesting reduced metastability of the higher energy E conformation. Taken together with the absence of soft terminal ligands in the anion layer to mediate cooperative motion, this finding warrants against glassy freezing in this material. 

For $\kappa$-CuNCS and $\kappa$-(CN)$_2$N(CN)$_2$, the lower symmetry of the $P2_1$ space group results in two inequivalent EEGs with the potential for thermal disorder. As shown in Fig. \ref{Fig8}(b), those in the A-sublattice possess close Z $\cdot\cdot\cdot$ H interactions with three neighbouring terminal ligands, while the B-sublattice EEGs are situated closer to the more rigid backbone Y ligands, and have close contacts with only one terminal SCN or N(CN)$_2$ group. In the case of the non-disordered $\kappa$-(CN)$_2$N(CN)$_2$, the A- and B-sublattice display large preference for the E and S conformation, respectively, with $2\Delta E/k_{\rm B}$ = 4530 and 2010 K. For the former A-sublattice, strong coupling between the EEG and thermally mobile Z = N(CN)$_2$ ligands results in significant thermal displacement parameters for the EEG, but no disorder arises, likely due to the strong stabilization of the eclipsed conformation. For the same sublattice in $\kappa$-CuNCS, strong van der Waals coupling to nearby Z = SCN ligands reverses this preference, favouring instead the staggered S-conformation, although with a much smaller $2\Delta E/k_{\rm B} = 1090$\,K. In this material, the EEGs in the B-sublattice were found to prefer the S conformation as well, although with a larger $2\Delta E/k_{\rm B}$ = 1360~K. The activation energies were also observed to slightly differ, being 3930 and 3790 K for the A- and B-sublattices, respectively. Thus, the complex two-stage glass transition in $\kappa$-CuNCS may be related to subsequent freezing of the two EEG sublattices existing in distinct local environments. In this scenario, close Z $\cdot \cdot \cdot$ H contacts may play a dual role, both mediating interactions within the A-sublattice, and linking the two sublattices together as shown in Fig. \ref{Fig8}(b). A strong coupling of the EEG and terminal ligand vibrational degrees of freedom is suggested by significant structural changes in the anion layer in the vicinity of 100 K and 60 K in $\kappa$-CuNCS, which preempt the glass transitions at $T_{g_1}$ = 70 K and $T_{g_2}$ = 53 K, respectively\cite{Doi1991}. These changes are particularly dramatic in modulation of the average Cu-N distance presumed to be due to motion of the terminal Z = SCN group. 

It has been previously suggested \cite{Hiramatsu2015,Yamochi1993} that the arrangement of the ET molecules and conformational preferences in $\kappa$-(ET)$_2$X salts with polymeric anions result from a particular pattern of empty spaces in the anion layers, allowing the donor molecules to fit in between, thus minimizing the van-der-Waals interactions. From this perspective, glassy freezing of the EEG rotation requires that the EEG-anion interaction is relatively ineffective at distinguishing the S and E states, allowing for similar energies of both states, and thus metastability.
Indeed, in the model described in Ref.\ \onlinecite{Hartmann2014} it can be shown that the degree of frozen-in disorder 
is determined by the ratio of $E_A/\Delta E$ (at fixed $\nu_0$), with increasing occupancy of the minor conformation below $T_g$ with increasing ratio. In this regard it is interesting to note that only those salts found to have $E_A/\Delta E$ above an empirical threshhold (here found computationally to be $\sim 5$) display an observable glass transition.

\section{Summary and Conclusion}
In summary, we present a comprehensive study of the glassy structural ordering of the ET molecules' ethylene endgroups in $\kappa$-(ET)$_2$X with polymeric anions X, a phenomenon which has important consequences for the properties of the strongly-correlated electron system. Due to the coupling of the thermal EEG vibrations to resistance fluctuations, the low-frequency dynamics of the glassy relaxation has been determined yielding the activation energy of the process in samples of $\kappa$-Br and $\kappa$-Cl in a model of non-exponential kinetics. The characteristic Vogel-Fulcher-Tammann behavior of the relaxation reveals that the present materials are fragile rather than strong orientational glass formers. {\it Ab initio} calculations emphasize that the coupling between the anion layer and the EEG degrees of freedom plays the decisive role if a glass transition occurs and that the collective motion of EEG and anions freezes out at $T_g$. The estimated parameters for the activation energies and energy differences between the E and S configurations explain why a double transition occurs for $\kappa$-CuNCS but no transition for $\kappa$-Cu$_2$(CN)$_3$. Predictions for the occurrence of glass-like transitions in certain (ET)$_2$X materials are now feasible.

\section*{Acknowledgments}
Work is supported by the Deutsche Forschungsgemeinschaft (DFG) within SFB/TR 49. S.M.W.\ thanks NSERC Canada for a postdoctoral fellowship. 
J.A.S.\ acknowledges support from the Independent Research/Development program at the National Science Foundation.
J.M.\ is grateful to Peter Lunkenheimer for valuable hints on the physics of glasses. 



\begin{thebibliography}{99}
\bibitem{comment2}
As can be seen from Fig.\,\ref{DDH_Daten}(a), the exponent cannot be determined unambiguously within the experimental error. Likewise, the temperature range, in which a constant offset in $\alpha(T)$ is observed, can somewhat vary for different systems and may be slightly sample dependent. $b = -3/2$ frequently gives the best possible description.
\bibitem{comment3}
We note that value of the maximum slightly depends on the correction related to the function $g(T)$ in the generalized DDH model, i.e.\ the coupling between the EEG fluctuations and the conduction electron system.
\end{thebibliography}

\end{document}